# β- $Cu_3V_2O_8$: Magnetic ordering in a spin-½ kagomé-staircase lattice

N. Rogado[1], M. K. Haas[1], G. Lawes[2], D. A. Huse[3], A. P. Ramirez[2], and R. J. Cava[1]

[1] *Department of Chemistry and Princeton Materials Institute, Princeton University, Princeton, New Jersey 08544*

[2] *Los Alamos National Laboratory, Los Alamos, New Mexico 87544*

[3] *Department of Physics, Princeton University, Princeton, New Jersey 08544*

E-mail: nsrogado@princeton.edu

**Abstract**
The spin-½ $Cu^{2+}$ ions in β-$Cu_3V_2O_8$ occupy the sites of a Kagomé-staircase lattice, an anisotropic variant of the Kagomé net: buckled layers and imbedded plaquettes of three edge-shared $CuO_4$ squares break the ideal Kagomé symmetry. Susceptibility and heat capacity measurements show the onset of short-range ordering at approximately 75 K, and a magnetic phase transition with the characteristics of antiferromagnetism at ~29 K. Comparison to the Curie Weiss theta ($\theta_{CW} = -135$ K) indicates that the geometric frustration is largely relieved by the anisotropy. A ferromagnetic contribution to the magnetization below the ordering temperature and negative magnetization in zero-field cooled measurements at low fields are attributed to uncompensated spins at grain boundaries or defects.

## 1. Introduction

Geometrically frustrated spin systems are based on the placement of antiferromagnetically interacting spins at the corners of a triangular lattice [1]. Such systems often remain magnetically disordered even when cooled well below the ordering temperature expected in a mean-field picture. For the two-dimensional (2D) triangular lattice of Heisenberg spins, the magnetic energy of an antiferromagnetic (AFM) system is reduced by adopting the 120° spin state. For the Kagomé geometry, a 2D lattice consisting of corner-sharing equilateral triangles, propagation of the 120° spin state throughout the lattice results in a large ground state degeneracy and frustration of long-range order [2]. The 2D Kagomé net is realized in many spin-1, spin-3/2, and spin-5/2 compounds exhibiting geometric magnetic frustration, including $SrCr_8Ga_4O_{19}$ [3,4], $Ba_2Sn_2Ga_3ZnCr_7O_{22}$ [5], and the jarosites $KM_3(OH)_6(SO_4)_2$ [6–8] with M = V, Cr, or Fe.

Theoretical studies have shown that magnetic order is not expected in the spin-½ Heisenberg antiferromagnet on the Kagomé lattice [9–11]. Recently Volborthite, $Cu_3V_2O_7(OH)_2 \cdot 2H_2O$ [12], a S = ½ Kagomé-like material, was reported. This compound consists of isosceles spin triangles instead of the ideal equilateral triangles, but magnetic measurements showed neither long-range ordering nor a spin-gap down to 1.8 K in spite of strong AFM interactions ($J/k_B = -84$ K and $\theta_{CW} \sim -100$ K). Here we report the magnetic characterization of β–$Cu_3V_2O_8$, a S = ½ magnetic material based on a novel anisotropic variation of the Kagomé net: the Kagomé-staircase lattice. The Kagomé-staircase is found in $M_3V_2O_8$ compounds with M = $Ni^{2+}$ (S = 1) [13], $Co^{2+}$ (S = 3/2) [14], $Zn^{2+}$ (S = 0) [15] and $Cu^{2+}$ (S = ½) [16]. These materials consist of buckled Kagomé layers of edge-sharing transition metal (II) oxide octahedra separated by nonmagnetic $V^{5+}O_4$ tetrahedra. The lower symmetry of the "staircase" magnetic layers and the resulting anisotropic intralayer superexchange interactions result in the reduction of geometric frustration and magnetic ordering in the Ni and Co variants [17]. Long range magnetic ordering in Kagomé antiferromagnets that have been doped with nonmagnetic impurities has been attributed to "order by disorder" that selectively pins down a particular spin configuration among the many possible ground states [18]. In contrast, magnetic ordering in the Kagomé-staircase materials is a consequence of breaking the ideal Kagomé symmetry in a periodic way through buckling of the magnetic layers. Spins with a smaller quantum number have a stronger tendency toward quantum disorder. Thus, the spin-½ Kagomé-staircase compound β–$Cu_3V_2O_8$ presents a unique opportunity to observe the competition between the frustrating influence of quantum effects and geometry versus the tendency toward magnetic ordering due to strong anisotropy.

## 2. Experiment

$\beta-Cu_3V_2O_8$ [16] was synthesized by heating $\alpha-Cu_3V_2O_8$ [19] in a cubic anvil press at 900 °C under a pressure of approximately 40 KBar for two hours. $\alpha-Cu_3V_2O_8$ was made from stoichiometric mixtures of CuO and $V_2O_5$ heated at 700 °C overnight in air. Nonmagnetic, isostructural $Zn_3V_2O_8$ was synthesized by heating ZnO, $V_2O_5$, and $ZnCl_2$ mixed in a 3:1:2 ratio at 600 °C for 4 hours in air. The product was washed with distilled water to remove the $ZnCl_2$ flux. Measurements on $Zn_3V_2O_8$ provide an estimate of the nonmagnetic contribution to the low temperature heat capacity of $\beta-Cu_3V_2O_8$.

## 3. Results and discussion

All of the samples were determined to be single-phase by powder X-ray diffraction. $\beta-Cu_3V_2O_8$ was found to be monoclinic with $a$ = 6.253(1) Å, $b$ = 8.002(1) Å, $c$ = 6.377(1) Å and $\beta$ = 111.45(2)°, in close agreement with previously reported data [16]. Unlike in the Ni and Co variants, the $Cu^{2+}O_6$ octahedra in $\beta-Cu_3V_2O_8$ are distorted significantly, with two longer apical Cu-O bonds, and four shorter in-plane bonds. The arrangement of the long and short bonds leads to the presence of plaquettes made of three edge-shared $CuO_4$ squares running nearly parallel to the $b$-direction within the basic Kagome net (figure 1a). The edge-sharing within the plaquettes leads to Cu-O-Cu angles between 85 and 95 degrees. The plaquettes are coupled through the apical oxygens, in a stretched edge-sharing geometry, with substantially longer Cu-O-Cu distances and near 90-degree Cu-O-Cu angles (figure 1b). The Cu site in the middle of the plaquette (Cu1) is crystallographically inequivalent to the other two (Cu2). The significant differences in the $Cu^{2+}$-O-$Cu^{2+}$ bond angles, and their divergence from the ideal edge-shared 90° angle, are all indicative of an anisotropic superexchange interaction. Magnetic coupling between the Kagomé layers is expected to be small based on the relatively large interlayer distance (~6.2 Å). Thus, $\beta-Cu_3V_2O_8$ is a 2D magnetic system dominated by anisotropic coupling interactions between the S = ½ spins on a Kagomé-based lattice.

Magnetic susceptibility data were obtained for $\beta-Cu_3V_2O_8$ between 2 and 300 K using a Quantum Design PPMS magnetometer. The field cooled (FC) magnetic susceptibility $\chi(T)$ of $\beta-Cu_3V_2O_8$ at an applied field of 1 Tesla is shown in figure 2, with its inverse susceptibility, $1/\chi$, as an inset. The high temperature $1/\chi$ data, above 200 K, was fitted to the Curie-Weiss law. The fit yielded a Curie constant $C$ = 0.57 emu/K mol Cu and a Curie-Weiss temperature $\theta_{CW}$ = −135 K. The effective magnetic moment ($\mu_{eff}$) for Cu was found to be 2.1 $\mu_B$, close to the expected moment for $Cu^{2+}$. The negative $\theta_{CW}$ value implies that AFM near neighbor interactions dominate. From the mean field theory, with $\theta_{CW} = zJS(S+1)/(3k_B)$, the superexchange interaction $J$ between $z$ nearest neighbor $Cu^{2+}$ (S = ½) ions on a Kagomé lattice is approximated by $\theta_{CW}$ [20]. The experimentally obtained $J/k_B$ value of −135 K is comparable to those found in other low-dimensional copper oxide compounds characterized by edge-sharing Cu-O-Cu bonds, such as $CuGeO_3$ [21]. Another estimate of the superexchange constant obtained by fitting the $\chi(T)$ data above 200 K to the high-temperature series expansion for a spin-½ Heisenberg antiferromagnet on an ideal Kagomé lattice [22] yielded $J/k_B$ = −90.3 K. The distortion of the Kagomé layers in $\beta-Cu_3V_2O_8$, however, limits the usefulness of this fit.

$\chi(T)$ shows a broad maximum at ~75 K, characteristic of the onset of short-range or low dimensional ordering. This may be attributed to the low-dimensional AFM ordering within the Kagome layers, or correlations within the embedded plaquettes of edge-shared Cu-O squares. A sharp feature appears at ~29 K, suggesting a three dimensional magnetic phase transition. The increase in $\chi(T)$ below this temperature is due to the contribution of a small amount of ferromagnetism accompanying the magnetic ordering.

Figure 3 shows the field dependence of the magnetization of $\beta-Cu_3V_2O_8$ at 5 K. Magnetic hysteresis is seen at lower fields. The loop at 5 K indicates a remnant magnetization of ~0.1 $\mu_B$, a small moment compared to the magnetization expected for a fully ferromagnetic (FM) spin-½ system. This FM contribution is observed to vary in different preparations of $\beta-Cu_3V_2O_8$, and is assumed to be an extrinsic effect. It is attributed to uncompensated AFM spins at grain boundaries, defects, or surfaces in the polycrystalline sample, as has been observed in thin films of layered antiferromagnets [23]. The intrinsic paramagnetic contribution to the susceptibility of $\beta-Cu_3V_2O_8$ below 29 K was obtained by measurement of M vs. H curves at different temperatures and deriving $\Delta M/\Delta H$ from the reversible region of the hysteresis loop. These data, which show that $\chi(T)$ drops significantly below 29 K, are shown in figure 2, confirming that the magnetic transition is due to AFM ordering.

Measurement of the M-H curve at small applied fields was obtained by setting the field to zero at 100 K, then cooling to 5 K, and then measuring the magnetization with increasing field starting from zero field. For the measurement shown in the inset of figure 3, the sample exhibits a diamagnetic moment below 400 Oe, although the slope of the M-H curve is positive. To investigate this further, zero-field cooled (ZFC) and field cooled (FC) measurements were taken at 0.1 Tesla (figure 4). Hysteretic behavior below 29 K is evident between the ZFC and FC measurements. The susceptibility of the FC measurement is greater than that of the ZFC measurement. The ZFC data is weakly positive at 5K, becomes negative, and then shows a sharp increase to match the field cooled data just below the magnetic transition. (It is noteworthy that the applied field for the ZFC data is more than twice that indicated by the M-H curve in figure 3 for a diamagnetic signal to persist in this sample.) Sign reversal of the net magnetization in ferromagnets such as spinels has been attributed to the different temperature and field dependences of the magnetization of the individual magnetic sublattices [24,25]. Although similar behavior is observed in $\beta-Cu_3V_2O_8$, the behavior of this sample is more consistent with uncompensated interfacial ferromagnetically aligned spins oriented by small remnant fields in the magnetometer during "zero-field cooling" rather than a bulk effect. The measurement of a negative magnetization is then just related to the relative orientations of the oriented spins at the interfaces and the measurement field.

The heat capacity was measured using a standard semi-adiabatic heat pulse technique. The magnetic contribution to the heat capacity ($C_m$) for $\beta-Cu_3V_2O_8$ was obtained by subtracting the lattice contribution from the raw data (figure 5). The lattice contribution was estimated from the heat capacity data of $Zn_3V_2O_8$, the nonmagnetic isostructural analogue of the Kagomé-staircase structure. The inset in figure 5 shows the raw data for $\beta-Cu_3V_2O_8$ and nonmagnetic $Zn_3V_2O_8$. $C_m$ for $\beta-Cu_3V_2O_8$ shows a sharp transition at 29 K, which is preceded by a broad feature at ~50 K. The observed transitions are consistent with the $\chi(T)$ data, that is, 3D AFM ordering preceded by low dimensional spin ordering. Integrating $C_m/T$ against T up to 100 K (assuming a smooth curve from 2 to 0 K) gives an entropy of ~5.6 J/mol Cu K, approximately equal to the entropy expected to fully disorder a mole of a spin-½ system.

## 4. Summary and conclusion

Based on this preliminary report of its magnetic properties, $\beta-Cu_3V_2O_8$ is shown to be an interesting magnetic system. A single 3D magnetic ordering transition is seen above 29 K, with indications of short-range or low dimensional ordering at higher temperatures. In contrast, the spin-1 analogue $Ni_3V_2O_8$ shows *four* magnetic transitions above 2 K and the spin-3/2 analogue $Co_3V_2O_8$ shows two magnetic transitions (and an incomplete accounting of the spin entropy in those transitions) [17]. Most theoretical and experimental work has focussed on ideal Kagomé lattices. Future studies on the novel Kagomé-staircase structure will clarify the effects of anisotropic magnetic coupling in the frustrated Kagomé net.

**Acknowledgments**
This research was supported by the National Science Foundation through the MRSEC program (NSF MRSEC grant DMR-9809483).

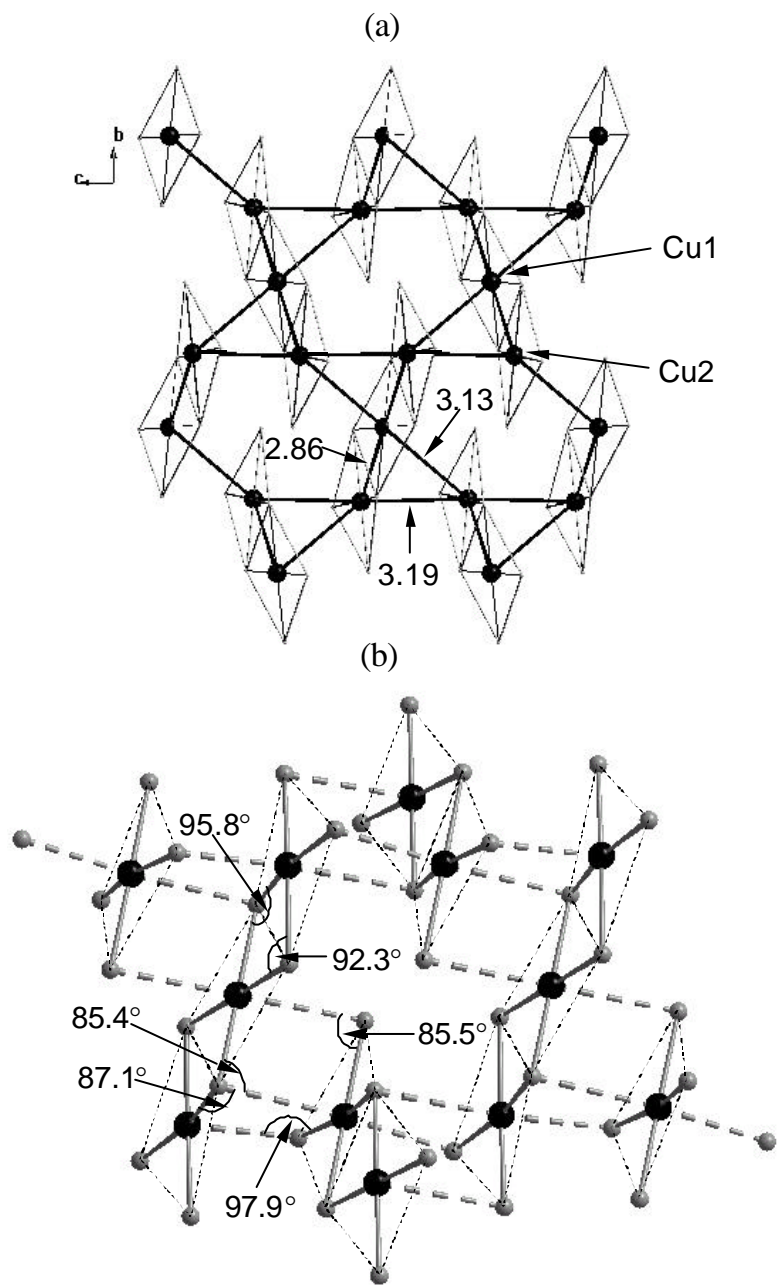

Figure 1. (a) The "Kagomé-staircase" lattice in $\beta-Cu_3V_2O_8$. The basic Kagome net is emphasized with bold black lines. Plaquettes of three edge-shared $Cu^{2+}O_4$ squares are shown. The inequivalent Cu sites are labelled, as are the Cu-Cu distances (in Å). (b) Detail of the local geometry around the Cu atoms, showing the Cu-O-Cu bond angles and the triple $CuO_4$ plaquettes. Black circles: Cu, gray circles: O. Solid gray lines are Cu-O bonds $\leq 2$ Å, broken gray lines are long Cu-O bonds ($\geq 2.3$ Å).

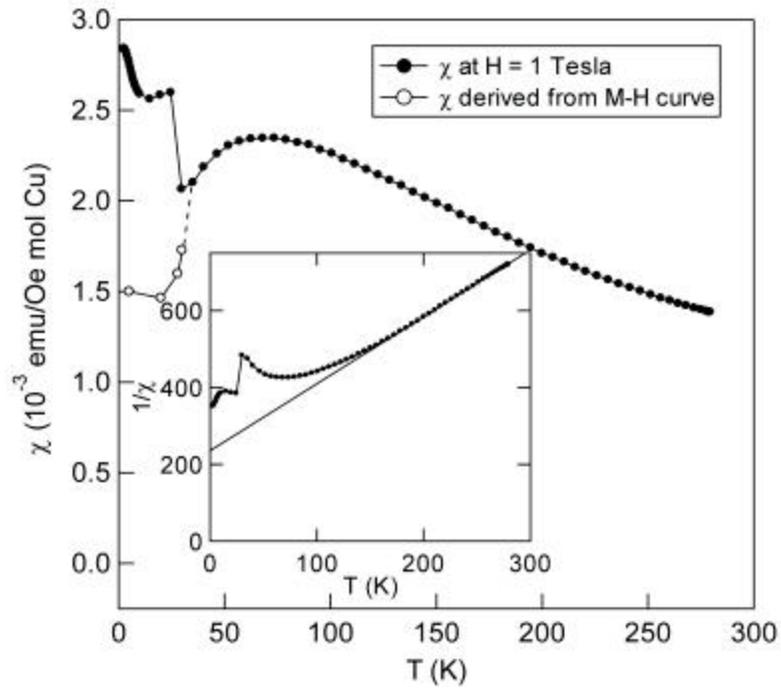

Figure 2. Field-cooled magnetic susceptibility of β−$Cu_3V_2O_8$ taken at H = 1 Tesla (solid circles) and the paramagnetic contribution to the susceptibility derived from M-H curves taken between 5 and 30 K (open circles). Inset shows inverse susceptibility, 1/χ. The solid line is the Curie-Weiss fit.

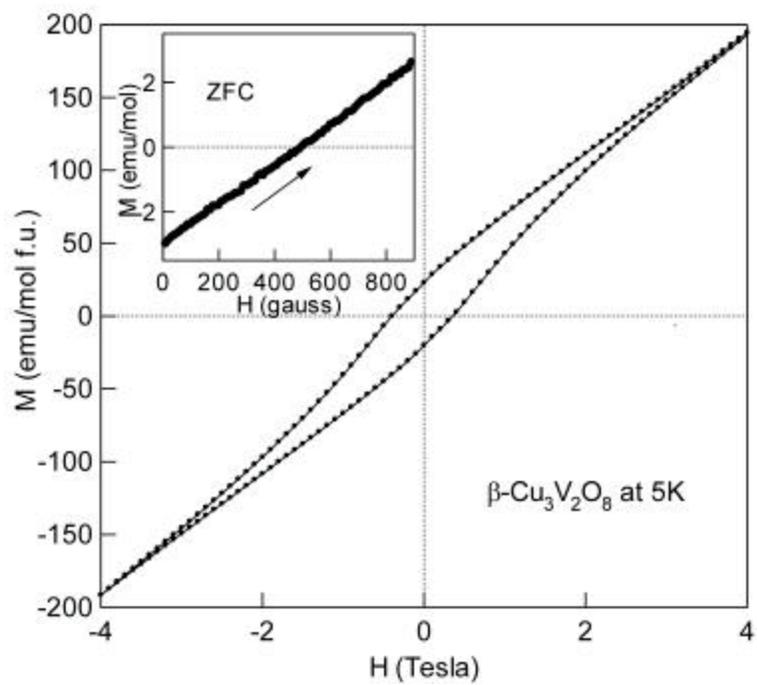

Figure 3. Magnetization vs. applied field at 5 K for $\beta-Cu_3V_2O_8$. Inset shows M/H plot at low fields.

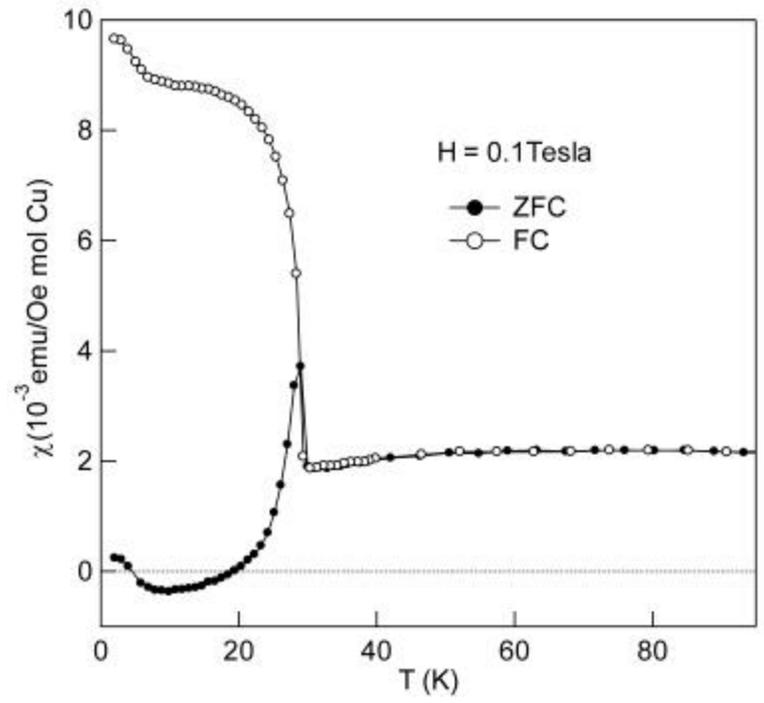

Figure 4. Zero-field cooled (ZFC) and field cooled (FC) measurements taken at 0.1 Tesla on heating.

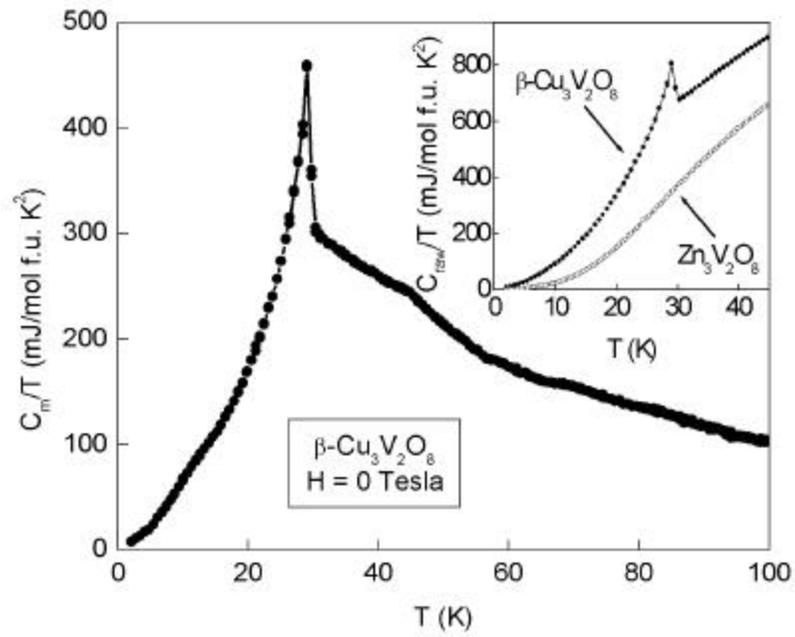

Figure 5. Temperature dependence of the magnetic contribution to the heat capacity of β−$Cu_3V_2O_8$ represented as $C_m/T$. Inset shows the raw data ($C_{raw}/T$) for β−$Cu_3V_2O_8$ and nonmagnetic $Zn_3V_2O_8$.